\documentstyle[12pt,fleqn]{article}
\begin{document}
\baselineskip=7.0mm

\begin{center}
{\large\bf A Two-Step Model for Gamma-Ray Bursts Associated with
Supernovae}

\vspace{5.0mm}
K. S. Cheng$^1$ and Z. G. Dai$^2$ 

$^1${\em Department of Physics, University of Hong Kong, Hong Kong, 
China}

$^2${\em Department of Astronomy, Nanjing University, Nanjing 210093,
China}

\end{center}

\vspace{3mm}

\begin{center}
ABSTRACT
\end{center}

We here propose a two-step model for gamma-ray bursts
(GRBs) associated with supernovae.
In the first step, the core collapse of a star with mass $\ge 19M_\odot$
leads to a massive neutron star and a normal supernova, and subsequently
hypercritical accretion of the neutron star from the supernova ejecta
may give rise to a jet through neutrino annihilation and/or  
Poynting flux along the stellar
rotation axis. However, because of too much surrounding matter, this
jet rapidly enters a nonrelativistic phase and evolves to a large 
bubble. In the second step, the neutron star promptly implodes to 
a rapidly rotating black hole surrounded by a torus when the mass of the 
star increases to the maximum mass and meanwhile its rotation frequency
increases to the upper limit due to the accreted angular momentum.
The gravitational binding energy of the torus may be dissipated by 
a magnetized relativistic wind, which may then be absorbed by the
supernova ejecta, thus producing an energetic hypernova. The rotational
energy of the black hole may be extracted by the Blandford-Znajek's
mechanism, leading to another jet. This jet is relatively free of 
baryons and thus may be accelerated to an ultrarelativistic phase
because the first jet has pushed out of its front matter and left 
a baryon-free exit. We expect that the second jet generates a GRB 
and its afterglow. Our two-step model may alleviate the
baryon-contamination problem suffered  possibly from in the 
hypernova models. Furthermore, this model not only accounts for
association of several GRBs with supernovae but also explains well 
the features of the afterglows of these bursts.          

\noindent
{\em Subject headings:} gamma-ray: bursts -- supernovae: general
-- stars: neutron -- black holes: physics

\vspace{2mm}

\begin{center}
1. INTRODUCTION
\end{center}

It has been widely believed that gamma-ray burst (GRB) 
events do indeed occur at cosmological distances (Metzger et al. 1997;
Kulkarni et al. 1998a, 1999; Andersen et al. 1999),
which implies that a successful model for progenitors 
of cosmological GRBs must satisfy two essential requirements: 
(1) The model should produce an extremely
relativistic fireball, which should subsequently emit an amount of 
gamma-ray isotropic energy $E_\gamma\sim 10^{51}$-$10^{54}\,$ergs 
implied by the observed fluences and the cosmological distance scale. 
The recent multi-wavelength observations of GRB afterglows support the
so-called fireball shock model (Piran 1999). 
According to this scenario, GRBs are produced 
as a result of internal shocks when fast moving shells
catch up with slower shells that were ejected at earlier times.
The relative kinetic energy of motion of the shells is converted into
observed gamma-ray emission in relativistic shocks via synchrotron
radiation or inverse-Compton scattering mechanisms. The observed
afterglow emission is produced when the shell decelerates as a result
of interaction with the ambient matter. Based on the fireball shock
model, several authors (Panaitescu, Spada \& M\'esz\'aros 1999;
Kumar 1999; Lazzati, Ghisellini \& Celotti 1999) recently found that
the efficiency for producing gamma-rays in internal shocks is a few
percent. Therefore, the isotropic energy of fireballs in some bursts
(e.g., GRB 990123) must be up to a few times $10^{55}\,$ergs
(Kulkarni et al. 1999). If anisotropic emission with a beaming factor 
of $\Delta\Omega/4\pi\sim 0.01$ is assumed, this energy can be reduced 
to $E_{\rm jet}\sim$ a few times $10^{53}\,$ergs. (2) The rapid
variability of GRBs and their nonthermal spectra (Woods \& Loeb 1995) 
and the low radiative efficiency in
internal shocks (Lazzati et al. 1999) requires that 
the Lorentz factor of the fireball be 100-1000. This implies that 
the fraction of contaminating baryons must be less than 1\%. If the
emission is anisotropic, the mass of loading matter $\Delta M\le
0.01E_{\rm jet}/c^2\sim 10^{-3}M_\odot$.

Two currently popular models of GRB progenitors are the merger of
two compact objects (neutron stars and black holes) and the
collapse of massive stars. The former model is the plausibly 
baryon-clean one, but compact objects would be expected to have
such a significant space velocity that their merger would take
place outside their birthplaces (Paczy\'nski 1998; Bloom, Sigurdsson
\& Pols 1999). The observational evidence for the association
of several GRBs with star forming regions then provided weak evidence
against the compact object merger as GRB progenitors and favored
massive star progenitors. The population synthesis of Fryer, Woosley
\& Hartmann (1999) supported this conclusion.

The massive star progenitor model has become more favorable since
the discovery of the Type Ic supernova SN 1998bw in the error box of
GRB 980425 (Galama et al. 1998). The high energy inferred for the optical
supernova, $(2-3)\times 10^{52}\,$ergs (Iwamoto et al. 1998; 
Woosley, Eastman \& Schmidt 1999), and the high expansion velocity 
inferred for the radio supernova (Kulkarni et al. 1998b) strengthen
the GRB-SN connection. Recently, this connection was further confirmed
by Bloom et al. (1999), Reichart (1999) and Galama et al. (1999), 
who found the dramatical brightening and extreme reddening 
of the optical afterglows of GRB 980326 and GRB 970228 
at late times, respectively. However, it is quite difficult
to understand such a connection. From the above second essential 
requirement, too many baryons possibly exist in the vicinity of the
collapsing core in the hypernova models (Woosley 1993; Paczy\'nski 1998) 
so that an ultrarelativistic jet forming during
the collapse of the core rapidly become nonrelativistic. This conclusion 
is consistent with the numerical studies (MacFadyen 
\& Woosley 1999): an ultrarelativistic jet converts to 
a nonrelativistic large bubble. A simple reason for this may be that a
large amount of radiative energy is released impulsively but the mass of
contaminating baryons is in fact of the order of
$M_*(\Delta\Omega/4\pi)\sim 0.1M_\odot
(M_*/10M_\odot)(10^2\Delta\Omega/4\pi)$, where $M_*$ is the mass
of the matter around the core.

In this Letter, we propose a scenario, in which a supernova
explosion first may produce a massive neutron star, and about two hours
later the star will start to accrete the fall-back supernova ejecta
at a hypercritical rate, while an accretion disk will form near 
the stellar surface due to its large angular momentum. In particular 
neutrino annihilation and/or Poynting flux along the rotation
axis of the star may lead to a jet which will pushes out of its front
matter. The mass of the accreting neutron star will eventually reach 
the maximum mass about several hours after the supernova explosion and
thus will promptly collapse to a rapidly rotating black hole surrounded 
by a torus. The gravitational binding energy of the torus, which is of the
order of several $10^{52}\,$ergs, may be dissipated into the supernova
ejecta, which may in turn give rise to an energetic hypernova.
Another jet with energy of a few times $10^{53}\,$ergs and with
low-mass baryon contamination will occur along the rotation axis of 
the black hole by extracting its rotational energy via the
Blandford-Znajek's (1977) mechanism. We expext that the second jet can
produce a GRB and its afterglow.

Delayed formation of black holes in supernovae has been widely 
discussed (Woosley 1988; Chevalier 1989; Brown \& Weingartner 1994;
Brown \& Bethe 1994; Woosley \& Weaver 1995; Fryer 1999). Recently, it
has been shown numerically that the accretion of such a black hole may
lead to a relativistic jet required by a typical GRB (Woosley, MacFadyen 
\& Heger 1999; MacFadyen, Woosley \& Heger 1999). The present
model is an analytical one, in which we suggest that hypercritical
accretion of a newborn neutron star and delayed formation of a black
hole could produce two discrete jets. The first jet will push out of its
front matter and leave an exit for the second jet, which could thus
be relatively free of baryons.   
        
\begin{center}
2. HYPERCRITICAL ACCRETION OF A NEWBORN NEUTRON STAR AND FORMATION
OF A BUBBLE
\end{center}

It is well known that the core collapse of massive stars with 
10-25$M_\odot$ produces neutron stars accompanying Type II
supernovae. Timmes et al. (1996) numerically studied the initial mass
function of newborn neutron stars and found that their initial mass
distribution is bimodal with peaks at $1.27$ and $1.76M_\odot$. The
principal reason for this bimodal distribution is the differnce in the
presupernova structure of stars above and below $19M_\odot$, the mass
seperating stars that burn carbon convectively from those that produce
less carbon and burn radiatively. Here we consider neutron stars with
$1.76M_\odot$ as the startimg point of our work.
When such a massive neutron star first forms in a supernova explosion, 
it is surrounded by a dense gas (supernova ejecta), some of 
which falls onto the neutron star and cools by neutrino emission
(Colgate 1971; Zeldovich, Ivanova \& Nad\"ezhin 1971). It is the 
neutrino emission that allows accretion of the star at a high rate.  
From simple analytical arguments, Chevalier (1989) and Brown \&
Weingartner (1994) estimated a lower limit to steady neutron star
accretion with neutrino losses assuming spherical symmetry. However,
the accreted matter may have a large angular momentum which leads 
to an accretion disk. In this case, the lower limit with neutrino
losses is estimated as $\dot{M}_{\rm cr}\sim 
1.1\times 10^{-3}M_\odot\,{\rm yr}^{-1}$
(Chevalier 1996). 

The supernova explosion scenarios involve an outgoing shock wave.
When this shock enters the hydrogen envelope, the
deceleration of matter occurs. This deceleration sharpens into
a reverse shock. When the reverse shock reaches the neutron star
surface, the star starts to accrete the fall-back supernova ejecta  
at a hypercritical rate. The time scale for the reverse shock to
pass through the core is about the same as the time that the outgoing
shock front takes to reach the stellar surface, which is 
$t_0\sim 2\,{\rm hr}(R_*/3\times 10^{12}{\rm cm})
(M_{\rm ej}/10M_\odot)^{1/2}(E_{\rm sn}/10^{51}{\rm ergs})^{-1/2}$,
where $R_*$ is the presupernova stellar radius, $M_{\rm ej}$ is the
mass of the ejected matter and $E_{\rm sn}$ is the supernova explosive 
energy (Shigeyama, Nomoto \& Hashimoto 1988). For SN1987A, $t_0$ is 
about 2 hr (Shigeyama et al. 1988). In order to derive the rate of
mass accretion onto the neutron star after $t_0$, we follow Brown
\& Weingartner (1994). By assuming the neutron star to be at rest
with respect to its ambient matter and using the Bondi's (1955) spherical
accretion theory, we obtain the accretion rate
\begin{equation}
\dot{M}=5.63\times 10^{-4}\left(\frac{M}{1.8M_\odot}\right)^2
\left(\frac{v_f}{10^8{\rm cm}\,{\rm s}^{-1}}\right)^{-15/8}
\left(\frac{t}{1{\rm yr}}\right)^{-15/8}M_\odot\,{\rm yr}^{-1},
\end{equation}
where $M$ is the neutron star mass ($\sim 1.8M_\odot$) and $v_f$
is the final velocity (after being slowed down by the reverse shock)
of the carbon-oxygen core. For SN1987A, $v_f\sim 6\times 10^7\,{\rm cm}\,
{\rm s}^{-1}$ (Woosley 1988). In deriving equation (1), we have assumed
that the ambient matter of the neutron star is radiation-dominated 
due to the effect of the outgoing shock. Owing to this assumption, 
our accretion rate is slightly larger than that of Brown \& Weingartner
(1994). The time at which radiation significantly affects
accretion can be estimated based on $\dot{M}(t_{\rm cr})=
\dot{M}_{\rm cr}$. Hence, we obtain this timescale
$t_{\rm cr}=0.7v_{f,8}^{-1}\,{\rm yr}$,
where $v_{f,8}=v_f/10^8{\rm cm}\,{\rm s}^{-1}$. Below, we take
$v_{f,8}=1$. When $t<t_{\rm cr}$, the accretion is hypercritical 
and the total accreted baryon mass is given by
\begin{equation}
\Delta M_{\rm acc}=\int_{t_0}^t\dot{M}dt=1.81M_\odot v_{f,8}^{-15/8}
[(t_0/1{\rm hr})^{-7/8}-(t/1{\rm hr})^{-7/8}].
\end{equation}
For the modern realistic equation of state for neutron matter chosen in
the next section, an accreting neutron star with initial mass 
of $1.76M_\odot$ will collapse
to a rapidly rotating black hole when $\Delta M_{\rm acc}=0.55M_\odot$.

Before the collapse, the accreted matter forms a disk near the neutron
star because the accreted angular momentum may be up to 
$3^{1/2}R_sc$ per gram, where $R_s$ is the Schwarzchild radius
(Woosley \& Chevalier 1989). The temperature of the accretion disk
near the neutron star can be estimated by the following equation:
$\eta\dot{M}c^2={\dot{\epsilon}}_\nu\Delta\Omega_dR_{ns}^3)/4$,
where $\eta$ is the efficiency for the conversion of the
gravitational energy to heat ($\sim 0.1$), $R_{\rm ns}$ is the neutron
star radius ($\sim 10^6$ cm), $\Delta\Omega_d$ is the solid angle of
the disk, and ${\dot{\epsilon}}_\nu = 1.0\times 10^{25}(T/{\rm MeV})^9
\,{\rm erg}\,{\rm s}^{-1}\,{\rm cm}^{-3}$
is the neutrino pair energy production rate per unit volume (Dicus 1972). 
For typical values of these parameters (e.g., $\Delta\Omega_d
\sim 3$), we obtain $T\sim 6$ MeV. The total energy for neutrino losses
is approximated by
$E_\nu=GM\Delta M_{\rm acc}/R_{\rm ns}\sim 3\times 10^{53}\,{\rm ergs}$.
Since anisotropic neutrino emission takes place due to the effect of the
disk (Kluzniak 1998), neutrino annihilation along the rotation axis 
of the neutron star leads to a jet. By using the 
efficiency of neutrino annihilation, $\chi\sim 0.3\%$ (Goodman,
Dar \& Nussinov 1987; Kluzniak 1998), we obtain the energy of the jet
\begin{equation}
E_{{\rm jet},1}\sim E_\nu\chi\sim 10^{51}\,{\rm ergs}.
\end{equation}
It should be pointed out that this efficiency is much larger than
that of Popham, Woosley \& Fryer (1999), in which an accretion disk
surrounding a black hole is advection-dominated. In the present case,
however, matter is accreted onto the neutron star surface where 
the conversion of the gravitational energy to heat ($\eta$) is much more 
efficient. Since this thermal energy is released via neutrino emission,
the efficiency of neutrino annihilation should increase 
substantially. In addition to the neutrino annihilation mechanism,
there is another possible mechanism to produce a jet proposed by
Katz (1997), in which the magnetic field amplified by the differential 
rotation of the disk may result in a strong Poynting flux. The jet 
produced by these mechanisms will push its front baryonic matter whose
velocity is given by
$v_{\rm jet}=(2E_{{\rm jet},1}/\Delta M')^{1/2}
\sim 3\times 10^9(M_*/10M_\odot)^{-1/2}
(10^2\Delta\Omega/4\pi)^{-1/2}\,{\rm cm}\,{\rm s}^{-1}$,
where $\Delta M'=M_*(\Delta\Omega/4\pi)$ is the mass of baryons loading 
with the jet. It should be noted that 
this velocity is much larger than that of the outgoing
shock. As numerically studied by MacFadyen \& Woosley (1999), this
nonrelativistic jet will expand laterally, producing a bubble. 
When $t\sim 5$ hr, at which the collapse of the neutron star may take 
place as argued in the next section, the radius of the bubble is
estimated as
$R_b\sim v_{\rm jet}(t-t_0)\sim 3\times 10^{13}(M_*/
10M_\odot)^{-1/2}(10^2\Delta\Omega/4\pi)^{-1/2}\,{\rm cm}$,
which is much larger than the presupernova stellar radius $R_*$.
  
\begin{center}
3. COLLAPSE OF THE NEUTRON STAR AND GENERATION OF A GRB
\end{center}

It has been argued by Brown \& Bethe (1994) and Bethe \& Brown (1995) 
that hypercritical accretion of neutron stars in supernovae (e.g.,
SN1987A) may lead to collapse of the stars to low-mass black holes.
These authors considered a soft equation of state like kaon condensation
for neutron matter as the starting point of their discussions.
Here we adopt a more realistic modern equation of state named UV14+TNI
(Wiringa, Fiks \& Fabrocini 1988). This equation of state gives the
property of a nonrotating neutron star at the maximum mass:
maximum gravitational mass $M_{max}=1.84M_\odot$, corresponding
baryon mass $M_b=2.17M_\odot$, radius $R_{\rm ns}=9.51$ km and 
moment of initia $I=1.5\times 10^{45}\,{\rm g}\,{\rm cm}^2$ 
(Wiringa et al. 1988). According to this property, the maximum 
frequency of rotation for this neutron star equation of state
is given by 
$\Omega_{max}=7.84\times 10^3(M_{max}/M_\odot)^{1/2}
(R_{\rm ns}/10\,{\rm km})^{-3/2}{\rm s}^{-1}=
1.17\times 10^4\,{\rm s}^{-1}$ (Cook, Shapiro \& Teukolsky 1994).
The property of a rotating neutron star at this maximum frequency and 
at the allowable maximum mass is as follows (Cook et al. 1994):
maximum gravitational mass $M_{max}=2.19M_\odot$, corresponding 
baryon mass $M_b=2.55M_\odot$, radius $R_{\rm ns}=12.7$ km and
angular momentum $J=2.85\times 10^{49}\,{\rm erg}\,{\rm s}$ or 
the Kerr rotation parameter $a=cJ/GM_{max}^2=0.67$.
      
After having these properties, we discuss implications of hypercritical
accretion of a newborn neutron star with gravitational mass of
$1.76M_\odot$. First, hypercritical accretion may produce and maintain
a large bubble, as argued in the last section. 

Second, hypercritically accreted matter may rapidly submerge 
the magnetic field of the neutron star (Muslimov \& Page 1995). 
Once the accreted mass reaches $0.01M_\odot$, 
the buried magnetic field ohmically diffuses out after $\sim 10^8$ yr 
(Geppert, Page \& Zannias 1999). This implies that the magnetic field
of the neutron star could be always weak in the accretion timescale.

Third, hypercritical accretion may spin up the neutron star 
(Woosley \& Chevalier 1989). A $1.76M_\odot$ neutron star has the 
baryon mass of $2.0M_\odot$ and thus this star needs to accrete 
matter with mass of $0.55M_\odot$ to become a maximum rotating
neutron star with maximum mass. As suggested by Woosley \& Chevalier 
(1989), the accreted angular momentum from the mixed mantle and helium
core of the ejecta may be as large as $3^{1/2}R_sc$ per gram 
corresponding to $5.54\times 10^{48}\,{\rm erg}\,{\rm s}$ per 
$0.1M_\odot$ accreted mass. Thus, after accreting $0.55M_\odot$
mass, the neutron star can obtain the angular momentum of 
$3\times 10^{49}\,{\rm erg}\,{\rm s}$. But, the maximum angular 
momentum of a rapidly rotating neutron star at maximum mass is 
only $2.85\times 10^{49}\,{\rm erg}\,{\rm s}$. The remaining angular
momentum is $\Delta J=0.15\times 10^{49}\,{\rm erg}\,{\rm s}$. How is
$\Delta J$ dissipated? Fortunately, this remaining accreted 
angular momentum can be easily carried away by gravitational
radiation (Wagoner 1984). Owing to hypercritical accretion, therefore,
the mass of the neutron star not only reaches the maximum mass 
but it is spun up to the maximum rotation frequency. 

Finally, once the mass of the accreted matter reaches $0.55M_\odot$,
the neutron star will promptly collapse to a rapidly rotating black
hole. From equation (2), we can estimate the time at which the collapse
will occur: $t\sim 5.1$ hr. The resulting black hole has the rotational
energy 
\begin{equation}
E_{\rm rot}=f(a) M_{\rm BH}c^2\approx 3\times 10^{53}\,{\rm ergs},
\end{equation}
where $f(a)=1-\sqrt{(1+\sqrt{1-a^2})/2}\approx 0.067$ at $a\approx 0.67$,
and the mass of the black hole has been taken to be $2M_\odot$.
After the collapse, not all mass will be immediately accreted; the outmost
layers with a small fraction (a few percent) of the total mass, in fact,
have centrifugal accelerations beyong the local gravitational attraction
(Vietri \& Stella 1998), leading to a torus. In addition, a small
amount of accreted ejecta still stagnate in the torus. Thus, the total 
mass of the torus may be $M_t\ge 0.1M_\odot$. The presence of the
torus will give rise to the following two effects: (i) The binding energy
of the torus is
$E_b=GM_{\rm BH}M_t/R_t=3\times 10^{52}\,{\rm ergs}
(M_t/0.1M_\odot)(R_t/20\,{\rm km})^{-1}$,
where $R_t$ is the typical radius of the torus. Because the magnetic field 
of the torus may be amplified to $10^{15}$ G due to differential 
rotation in the torus, this binding energy can be dissipated into a 
magnetized relativistic wind (Usov 1994; M\'esz\'aros \& Rees 1997;
Katz 1997). Because such a wind may be easily absorbed by the 
outgoing supernova ejecta as argued by Dai \& Lu (1998a, 1998b), almost
all of the binding energy may convert to the expansion energy of the 
ejecta, which may further give rise to a supernova with a much 
brighter optical luminosity. 
(ii) In the presence of the torus, the rotational energy 
of the hole can be extracted by the Blandford-Znajek (1977) mechanism
(M\'esz\'aros, Rees \& Wijers 1998). The power for this mechanism
is $P_{\rm BZ} = 1.7\times 10^{51}a^2f(a)(M_{\rm BH}/M_\odot)^2
(B/3\times 10^{15}{\rm G})^2{\rm erg}\,{\rm s}^{-1}  
= 2.5\times 10^{50}(B/3\times 10^{15}{\rm G})^2{\rm erg}
\,{\rm s}^{-1}$ (Lee, Wijers \& Brown 1999).
The rotational energy of the black hole will be dissipated in
the timescale of $\sim 10^3$ s if the magnetic field strength is 
of the order of $3\times 10^{15}$ G. We believe that this 
mechanism will produce an ultrarelativistic jet because the mass
of matter loading with this jet can be estimated by
\begin{equation}
\Delta M=\Delta M'(\Delta\Omega/4\pi)=
10^{-3}M_\odot (M_*/10M_\odot)(10^2\Delta\Omega/4\pi)^2.
\end{equation}
In fact, since the angular momentum of the presupernova star may be  
rather large, the centrifugal force should reduce the density of matter
along the rotation axis and thus the above estimate is an upper
limit. After a short acceleration phase, the Lorentz factor of the jet
should be
\begin{equation}
\Gamma=E_{\rm rot}/(\Delta Mc^2)
\ge 150 (M_*/10M_\odot)^{-1}(10^2\Delta\Omega/4\pi)^{-2},
\end{equation}
which is consistent with the constraints from the rapid variability of 
the light curves and the nonthermal spectra of GRBs (Woods \& Loeb 1995).    
The collisions among the shells in such a jet will produce a GRB and 
subsequently the deceleration of the jet in its ambient medium will
result in an afterglow.

\begin{center}
4. DISCUSSIONS	
\end{center}

We have proposed a two-step model for GRBs associated with supernovae. 
A hypernova, a much brighter supernova, may occur in our model. 
However, this model is clearly different from the current 
hypernova models (Woosley 1993; Paczy\'nski 1998; MacFadyen \& Woosley
1999). In the latter models, the core of a massive star directly
collapses to a black hole. Recently, Cen (1998) and Wang \& Wheeler
(1998) proposed models for GRB-SN association, in which the matter 
above the neutrinosphere in a small cone around some special axis of 
a newborn neutron star is assumed to be preferentially first blown out of 
the deep gravitational potential well of the star in order to avoid too
many baryons contaminating a subsequently resulting jet. Our model may
provide a plausible way of how such an empty cone is produced: neutrinos
from the hypercritical accretion disk annihilate to electron/positron
pairs which form the first jet to push its front baryons and leave an exit
for the second jet. Therefore, our model may avelliate the baryon
contamination problem. 

In our model, GRBs are naturally associated with supernovae because 
the former phenomenon takes place several hours after the latter
phenomenon. This is consistent with the analysis of Iwamoto et al. 
(1998), in which the time of core collapse coincides with that of
SN1998bw within ($+0.7$, $-2.0$) days. The unusually large explosive
energy of SN1998bw, $(2-3)\times 10^{52}$ ergs (Iwamoto et al. 1998;
Woosley et al. 1999), is very close to the gravitational binding 
energy between the black hole and torus, $\sim 3\times 10^{52}$ ergs. 
This consistency seems to support our model. 

The features of some afterglows can also be explained in the context 
of our model. For example, the optical afterglow of GRB 980326
rapidly decayed as $\propto t^{-2.0\pm 0.1}$ in the first two days, 
subsequently brightened dramatically and reddened significantly, 
and finally declined (Bloom et al. 1999). These features were also  
seen in the light curve of the afterglow of GRB 970228 (Reichart 1999; 
Galama et al. 1999). The late-time afterglows were widely believed
to be the contribution of unusually brighter supernovae, while the rapid
decay of the early afterglows was analytically argued to be due to
sideways expansion of jets (Rhoads 1999; Sari, Piran \& Halpern 1999).
Another interpretation for the rapidly decaying afterglows 
may be that a relativistic shock expanding in a dense medium
has evolved to a nonrelativistic phase (Dai \& Lu 1999a, 1999b). 
In our two-step model, this dense medium may be a presupernova 
steller wind. 

In summary, our two-step model may alleviate the
baryon-contamination problem suffered possibly from in the 
hypernova models. This model not only accounts for association of 
several GRBs with supernovae but also explains well the features of 
the afterglows of these bursts. 

We would like to thank J. S. Bloom, B. Hansen, A. MacFadyen and S. E.
Woosley for their comments and discussions. This work was supported 
by a RGC grant of Hong Kong government and the National Natural Science
Foundation of China.

\baselineskip=4mm

\begin{center}
REFERENCES
\end{center}

\begin{description}
\item Andersen, M. I. et al. 1999, Science, 283, 2075
\item Bethe, H. A., \& Brown, G. E. 1995, ApJ, 445, L129
\item Blandford, R. D., \& Znajek, R. L. 1977, MNRAS, 179, 433
\item Bloom, J., Sigurdsson, S., \& Pols, O. 1999, MNRAS, 305, 763
\item Bloom, J. et al. 1999, Nature, 401, 453 
\item Brown, G. E., \& Weingartner, J. C. 1994, ApJ, 436, 843
\item Brown, G. E., \& Bethe, H. A. 1994, ApJ, 423, 659
\item Bondi, H. 1952, MNRAS, 112, 195
\item Cen, R. 1998, ApJ, 507, L131
\item Chevalier, R. A. 1989, ApJ, 346, 847
\item Chevalier, R. A. 1996, ApJ, 459, 322
\item Colgate, S. A. 1971, ApJ, 163, 221
\item Cook, G. B., Shapiro, S. L., \& Teukolsky, S. A. 1994, ApJ, 424, 823
\item Dai, Z. G., \& Lu, T. 1998a, Phys. Rev. Lett., 81, 4301
\item Dai, Z. G., \& Lu, T. 1998b, A\&A, 333, L87
\item Dai, Z. G., \& Lu, T. 1999a, ApJ, 519, L155
\item Dai, Z. G., \& Lu, T. 1999b, astro-ph/9906109
\item Dicus, D. A. 1972, Phys. Rev. D, 6, 941
\item Fryer, C. 1999, ApJ, 522, 413
\item Fryer, C., Woosley, S. E., \& Hartmann, D. H. 1999, ApJ, in press
        (astro-ph/9904122)
\item Galama, T. J. et al. 1998, Nature, 395, 670
\item Galama, T. J. et al. 1999, ApJ, submitted (astro-ph/9907264)
\item Geppert, U., Page, D., \& Zannias, T. A\&A, 345, 847
\item Goodman, J., Dar, A., \& Nussinov, S. 1987, ApJ, 314, L7
\item Iwamoto, K. et al. 1998, Nature, 395, 672
\item Lazzati, D., Ghisellini, G., \& Celotti, MNRAS, in press
        (astro-ph/9907070)
\item Lee, H. K., Wijers, R. A. M. J., \& Brown, G. E. 1999, 
        astro-ph/9905373
\item Katz, J. I. 1997, ApJ, 490, 633
\item Kluzniak, W. 1998, ApJ, 508, L29
\item Kulkarni, S. R. et al. 1998a, Nature, 395, 663
\item Kulkarni, S. R. et al. 1998b, Nature, 393, 35
\item Kulkarni, S. R. et al. 1999, Nature, 398, 389
\item Kumer, P. 1999, ApJ, in press (astro-ph/9907096)
\item MacFadyen, A., \& Woosley, S. E. 1999, ApJ, 524, 262
\item MacFadyen, A., Woosley, S. E., \& Heger, A. 1999, ApJ, submitted 
        (astro-ph/9910034)
\item M\'esz\'aros, P., \& Rees, M. J. 1997, ApJ, 482, L29
\item M\'esz\'aros, P., Rees, M. J., \& Wijers, R. A. M. J. 1998,
        ApJ, 499, 301
\item Metzger, M. et al. 1997, Nature, 387, 878
\item Muslimov, A. \& Page, D. 1995, ApJ, 440, L77
\item Paczy\'nski, B. 1998, ApJ, 494, L45
\item Panaitescu, A., Spada, M., \& M\'esz\'aros, P. 1999,
        astro-ph/9905026
\item Piran, T. 1999, Phys. Rep., 314, 575
\item Popham, R., Woosley, S. E., \& Fryer, C. 1999, ApJ, 518, 356
\item Reichart, D. E. 1999, ApJ, 521, L111
\item Rhoads, J. 1999, ApJ, in press (astro-ph/9903399)
\item Sari, R., Piran, T., \& Halpern, J. P. 1999, ApJ, 519, L17
\item Shigeyama, T. Nomoto, K., \& Hashimoto, M. 1988, A\&A, 196, 141
\item Timmes, F. X., Woosley, S. E., \& Weaver, T. A. 1996, ApJ, 457, 834
\item Usov, V. V. 1994, MNRAS, 267, 1035
\item Vietri, M., \& Stella, L. 1998, ApJ, 507, L45
\item Wagoner, R. V. 1984, ApJ, 278, 345
\item Wang, L., \& Wheeler, J. C. 1998, ApJ, 504, L87
\item Wiringa, R. B., Fiks, V., \& Fabrocini, A. 1988, Phys. Rev. C, 38,
        1010
\item Woods, E., \& Loeb, A. 1995, ApJ, 453, 583
\item Woosley, S. E. 1988, ApJ, 330, 218
\item Woosley, S. E. 1993, ApJ, 405, 273
\item Woosley, S. E., \& Chevalier, R. A. 1989, Nature, 338, 321
\item Woosley, S. E., Eastman, R. G., \& Schmidt, B. P. 1999, ApJ, 516,
        788
\item Woosley, S. E., MacFadyen, A., \& Heger, A. 1999, astro-ph/9909034
\item Woosley, S. E., \& Weaver, T. A. 1995, ApJS, 101, 181
\item Zel'dovich, Ya. B., Ivanova, L. N., \& Nad\"ezhin, D. K. 1972,
        Soviet Astron., 16, 209
\end{description}

\end{document}